\documentclass[12pt]{article}

\advance \topmargin by -\headheight
\advance \topmargin by -\headsep

\evensidemargin \oddsidemargin
\usepackage{amssymb}
\usepackage{latexsym}
\usepackage{amsmath}
\usepackage{graphicx}

\newcommand{\Ref}[1]{(\ref{#1})}
\newcommand{\Hil}{\ensuremath{\mathcal{H}}}
\newcommand{\HilA}{\ensuremath{\mathcal{H_A}}}
\newcommand{\bracket}[2]{\ensuremath{\left<#1\left.\vphantom{#1}\vphantom{#2}\right|#2\right>}}   
   
\newcommand{\Dom}{\ensuremath{\mathrm{Dom}}}
\newcommand{\del}{\partial}
\newcommand{\supp}{\ensuremath{\mathrm{supp}}}
\newcommand{\A}{\ensuremath{\mathcal{A}}}
\newcommand{\C}{\ensuremath{\mathcal{C}}}
\newcommand{\X}{\ensuremath{\mathcal{X}}}
\newcommand{\Y}{\ensuremath{\mathcal{Y}}}
\newcommand{\Z}{\ensuremath{\mathcal{Z}}}
\newcommand{\cipci}{\ensuremath{C_0^\infty(\Sigma)\oplus C_0^\infty(\Sigma)}}
\newcommand{\textpf}[1]{\textbf{#1}}
\newcommand{\h}[1]{\ensuremath{H^{#1}_\mathrm{{loc}}}}

\newtheorem{theorem}{Theorem}[section]
\newtheorem{proposition}{Proposition}[section]
\newcounter{condition}
\renewcommand{\thecondition}{\textrm{(\Alph{condition})}}

\title{Dynamics in Stationary, Non-Globally-Hyperbolic  Spacetimes}
\author{Itai Seggev\thanks{\tt
iseggev@uchicago.edu} \\ \it Enrico Fermi Institute and Department
of Physics \\ \it University of Chicago \\ \it 5640 S.~Ellis Avenue,
Chicago, IL~60637, USA}
\date{November 20, 2003}

\numberwithin{equation}{section}

\begin{document}
\maketitle
\begin{abstract}
Classically, the dynamics of a scalar field in a
non-globally-hyperbolic spacetime is ill posed. Previously, a
prescription was given for defining dynamics in static spacetimes in
terms of a second-order operator acting on a Hilbert space defined on
static slices. The present work extends this result by giving a
similar prescription for defining dynamics in stationary spacetimes
obeying certain mild assumptions. The prescription is defined in terms
of a first-order operator acting on a different Hilbert space from the
one used in the static prescription. It preserves the important
properties of the earlier prescription: the formal solution agrees with the
Cauchy evolution within the domain of dependence, and smooth data of
compact support always give rise to smooth solutions. In the static
case, the first-order formalism agrees with second-order formalism
(using specifically the Friedrichs extension). Applications to field
quantization are also discussed.
\end{abstract}

\section{Introduction}
Let $(M, g_{ab})$ be a spacetime, and consider the minimally coupled,
massive Klein-Gordon equation for a real scalar field\footnote{A
complex scalar field may be treated by analyzing its real and
imaginary parts separately.} in this spacetime:
\begin{equation}
\label{eq:kg}
(g^{ab}\nabla_a\nabla_b - m^2) \varphi = 0. 
\end{equation}
As is well known (see, e.g., \cite{otob}), if $(M, g_{ab})$ is
globally hyperbolic, the Klein-Gordon equation is well posed. That is,
given smooth data on a spatial (Cauchy) hypersurface $\Sigma_0$, there
exists a unique smooth solution $\varphi : M \rightarrow \mathbb{R}$
with the given initial data. In a non-globally-hyperbolic spacetime,
however, no similar result holds. It is possible that no solution
exists to given initial data, or, if solutions exist, there may be
many which satisfy the initial condition. On the other hand, an
analysis of Einstein's equation led Penrose to his ``strong'' cosmic
censorship conjecture \cite{strong_cosmic}, which postulates that all
``reasonable'' spacetimes are globally hyperbolic. For these reasons,
the study of dynamics has often been restricted to globally hyperbolic
spacetimes.

In the absence of strong evidence either for or against cosmic
censorship, Wald \cite{nghsdI} considered whether a sensible
prescription could be given to define dynamics in a
non-globally-hyperbolic spacetime. He studied stably causal, static
spacetimes, that is, spacetimes with a hypersurface orthogonal Killing
vector field, whose orbits are complete and everywhere
timelike.\footnote{The standard black hole solutions do not fall into
the class of spacetimes being studied because their ``time
translation'' vector field becomes spacelike inside the event
horizon.} In these spacetimes, the Klein-Gordon equation can be split
into its spatial and temporal parts, leading to a second-order
equation:
\begin{equation}
\label{eq:static_eq}
\frac{\del^2 \phi}{\del t^2} = - A \phi~~~~~~~~A := -\alpha
D^a(\alpha D_a) + \alpha^2m^2,
\end{equation}
where $\alpha$ denotes the norm of the Killing field and $D_a$ the
spatial covariant derivative.  The operator $A$ is a positive
symmetric operator on an appropriate Hilbert space of initial data.
Thus, at least one self-adjoint extension,\footnote{The theory of
self-adjoint extensions of a symmetric operator is reviewed in Section
\ref{sec:operators}.} the Friedrichs extension, always exists
\cite{reed_simonII}.  The standard calculus of self-adjoint operators
(see, e.g., \cite{reed_simonI, rudin}) can therefore be used to define a
solution
\begin{equation}
\label{eq:static_sol}
\phi_t = \cos \left(A_E^\frac{1}{2} t\right) \phi_0 +
A_E^{-\frac{1}{2}} \sin \left(A_E^\frac{1}{2} t\right) (\alpha \pi_0),
\end{equation}
where $A_E$ denotes some self-adjoint extension of $A$ (not
necessearily the Friedrichs extension $A_F$), $\phi_0$ the
initial field configuration, and $\pi_0 $ the initial canonical
momentum (the canonical momentum is $\pi = n^a\nabla_a \phi,$ where
$n_a$ denotes the unit normal to the static slices). For initial data
in $C_0^{\infty}$, this solution is smooth throughout spacetime,
agrees with the classical solution within the domain of dependence
$D(\Sigma_0)$, and solves the Klein-Gordon equation in the pointwise
(spacetime) sense. Thus, the map $(\phi_0, \pi_0)
\rightarrow \phi_t$ assigns to each initial datum pair a unique
solution with sensible properties, and thus gives rise to a reasonable
prescription in the sense defined by Wald and Ishibashi
\cite{nghsdII}. Indeed, \cite{nghsdII} establishes that any reasonable
prescription must correspond to the one above for some choice of
positive, self-adjoint extension $A_E$. In this sense, the choice of
extension encodes the ``boundary conditions at the singularity'' (or
whatever feature gives rise to the non-globally-hyperbolicity).

This paper is concerned with a similar prescription for stably casual,
stationary spacetimes, i.e., those which possess a complete 1-parameter
family of timelike symmetries whose orbits are not necessarily
hypersurface orthogonal. In this case, there are ``mixed space-time
derivatives'' in the Klein-Gordon equation, so the division into space
and time parts of the equation that was done in \Ref{eq:static_eq} is
no longer possible. However, following Kay \cite{kay} by rewriting the
Klein-Gordon equation in Hamiltonian form and using an ``energy-norm''
Hilbert space, the time-evolution operator can be made into a
skew-symmetric operator. When this operator can be extended to a
skew-adjoint operator, a prescription analogous to \Ref{eq:static_sol}
can be given.

In contrast to the static case, it is not known \textit{a priori} that
any skew-adjoint extension of the time-evolution operator
exists. However, if the Killing field does not approach a null vector
(in a precise sense explained further in Section
\ref{sec:prescription}), it is possible to show that skew-adjoint
extensions exist when fields with positive mass are considered. Under
these same conditions, this prescription preserves the important
properties of the second-order prescription: it agrees with the
differential equation within the domain of dependence, and smooth data
of compact support give rise to smooth solutions. Further, in the
static case it is possible to compare the first and second-order
formalisms and show that they agree so long as the Freidrichs
extension $A_F$ is used in \Ref{eq:static_sol}. An interesting
contrast between the first and second-order formalisms becomes
apparent during this analysis. In the second-order formalism, a single
Hilbert space can be used to define all reasonable dynamics, and the
``boundary conditions at the singularity'' are determined by the
choice of self-adjoint extension.  In order to handle different
boundary conditions, the first-order formalism must be modified to
allow for different Hilbert spaces. Within each one, there is only one
skew-adjoint extension of the time-evolution operator. Hence the
choice of Hilbert space, not of skew-adjoint extension,
determines the dynamics and thus encodes the boundary conditions.

This paper is organized as follows: Section \ref{sec:operators}
reviews the theory of self-adjoint extensions of symmetric operators
and proves 
a corresponding result regarding skew-adjoint extensions of
skew-symmetric operators. Section \ref{sec:prescription} introduces
the prescription in detail. In Section \ref{sec:properties}, it is
shown that skew-adjoint extensions of the time-evolution operator
exist. It is further shown that the prescription agrees with the
differential equation within the domain of dependence and that smooth
data of compact support give rise to smooth solutions.
Section \ref{sec:static} examines the static case and compares the
first and second-order formalisms. As defining the space of classical
solutions is the first step towards field quantization, Section
\ref{sec:quantum} show how this prescritption may be used to extend
the well-known constructions of globally hyperbolic spacetimes 
to non-globally-hyperbolic ones. Section
\ref{sec:conclusions} concludes with some open problems.

\section{Extensions of Operators}
\label{sec:operators}
Although it is common in the physics literature to treat symmetric
operators (also called Hermitian operators) as equivalent to
self-adjoint operators, these two classes are in fact distinct. The
key difference is that arbitrary functions (and, in particular, the
exponential) of self-adjoint operators may be defined, whereas this is
not true of a general symmetric operator. The reason for this
difference is that the proper definition of an operator involves not
only specifying its action on vectors but also its domain. A symmetric
operator $A$ on a Hilbert space $\Hil$ is a linear operator defined on
a dense vector subspace $\Dom\ A \subseteq \Hil$, which ``acts the
same to the left or to the right'' for all vectors in its domain:
$$\bracket{\chi}{A\psi} = \bracket{A\chi}{\psi}~ \forall ~\psi, \chi
\in \Dom\ A.$$ A self-adjoint operator is a symmetric operator whose
domain is equal to the domain of its adjoint. While $\Dom\ A$ may be
freely chosen provided it is a dense vector subspace of $\Hil$, the
domain of $A^*$ is fixed once $A$ is defined. The somewhat convoluted
definition is as follows: $u \in \Dom\ A^*$ if and only if $\exists v
\in \Hil$ such that $$\bracket{u}{A\psi} = \bracket{v}{\psi} ~\forall~
\psi \in \Dom~A,$$ in which case $A^*u=v.$ Notice that for a symmetric
operator, $\Dom\ A$ is automatically contained in $\Dom\ A^*$, so if
the domains are not equal then $\Dom\ A \subsetneq \Dom\ A^*$. The
equality of domains for self-adjoint operators is crucial to the proof
of the spectral theorem, which allows arbitrary (measurable) functions
of the operator to be defined. For a (closed,) symmetric, non-self-adjoint operator,
functions may only be defined by a power series, but this
series will always diverge unless it has a finite number of terms. 

The close relationship between symmetric and self-adjoint operators
suggests that a symmetric operator can be turned into self-adjoint
operator by enlarging its domain of definition. That this is often the
case for operators acting on a complex Hilbert space is a consequence
of a famous theorem of von Neumann \cite{neumann, reed_simonII}.
\begin{theorem}
\label{thm:neumann}
Let $\Hil$ be a Hilbert space over $\mathbb{C}$ and $A:
\Hil \rightarrow \Hil$ a symmetric operator. Let $K_\pm :=
\mathrm{Ker} \left(A^* \mp i\right)$, and define $n_{\pm} :=
\mathrm{Dim}\ K_\pm$.  Self-adjoint extensions of $A$ exist if and
only if \mbox{$n_+=n_-=:n$}, in which case they are parametrized by
the group $U(n)$.
\end{theorem}
The indices $n_{\pm}$ are called the deficiency indices. If
\mbox{$n_{\pm}=0$}, there is a unique self-adjoint extension and $A$
is called essentially self-adjoint.  A standard technique for proving
that the deficiency indices are equal is the use of a complex
conjugation operator. Any involution (i.e., an operator whose square
is the identity) $C$ is called a complex conjugation if it is
antilinear and norm preserving.  Suppose that $\chi_{\pm} \in \Hil$
obeys $A^* \chi_\pm = \pm i \chi_\pm$. Assume further that $\exists$
some complex conjugation operator $C$ which commutes with $A$. For any
operator $A$, not necessarily symmetric, $[A, C] = 0$ implies that
$[A^*,C] = 0$ \cite[p. 360]{stone}. Thus, it follows that $A^*
C\chi_\pm = \mp i C\chi_\pm$; that is, $C$ establishes an isomorphism
between $K_+$ and $K_-$ so that the deficiency indices are equal. Von
Neumann also proved an extension of this result \cite{neumann,stone}.
\begin{theorem}
\label{thm:commute}
Let $\Hil$ be a Hilbert space over $\mathbb{C}$
and $A: \Hil \rightarrow \Hil$ a symmetric operator. If $\ \exists$ a
complex conjugation $C: \Hil \rightarrow \Hil$ with $[A, C]=0$, then
(i) $A$ has self-adjoint extensions, and (ii) $\exists$ self-adjoint
extensions of $A$ with $C$-invariant domain.
\end{theorem}

The previous two theorems only apply to complex Hilbert
spaces. However, the theory of extensions on a real Hilbert space is
naturally related to the theory on complex spaces by the following
proposition.
\begin{proposition}
\label{prop:sym_extend}
Let $\Hil_\mathbb{R}$ be a Hilbert space over $\mathbb{R}$, $A:
\Hil_\mathbb{R} \rightarrow \Hil_\mathbb{R}$ a symmetric operator, and
$A^\mathbb{C}$ the corresponding operator acting on $\Hil_{\mathbb{R}}
\otimes \mathbb{C}.$ The self-adjoint extensions of $A$ are in
1-1 correspondence with those self-adjoint extensions of
$A^\mathbb{C}$ whose domain is invariant under $\C$, the natural
complex conjugation operator defined on $\Hil_\mathbb{R} \otimes \mathbb{C}$.
\end{proposition}
This result is implicit in the original work of von Neumann and Stone
\cite[Theorems 8.1 and 9.14]{stone}.  Since $A^\mathbb{C}$ obviously
commutes with $\mathcal{C}$, parts (i) and (ii) of Theorem
\ref{thm:commute} combined with Proposition \ref{prop:sym_extend}
trivially show:
\begin{proposition}
\label{prop:sym_exist}
Let $\Hil_\mathbb{R}$ be a Hilbert space over $\mathbb{R}$. Any
symmetric operator  \mbox{$A: \Hil_\mathbb{R} \rightarrow
  \Hil_\mathbb{R}$} has self-adjoint extensions. 
\end{proposition}
In his original work \cite{nghsdI}, Wald used the positivity of the
operator $A$ in \Ref{eq:static_eq} to assert the existence of
self-adjoint extensions. However, Proposition \ref{prop:sym_exist}
shows that this was unnecessary. Existence follows directly from the
fact that $A$ is a symmetric operator defined on a real Hilbert space.

The operator considered in the present work is not symmetric but
rather skew-symmetric (or anti-Hermitian). Skew-symmetric operators
share many properties of symmetric operators, but pick up a minus sign
when the operator is transposed from the bra to the ket (or conversely):
$$\bracket{\chi}{A\psi} = -\bracket{A\chi}{\psi}~ \forall ~\psi, \chi
\in \Dom\ A.$$ In the case of a complex Hilbert space, there is a
one-to-one correspondence between symmetric and skew-symmetric
operators because multiplication by $i$ turns one type of operator
into the other. Thus, the analogue of Theorem \ref{thm:neumann} is
trivial. In the real Hilbert space case there is no such
correspondence because multiplication by $i$ is not a well-defined
operation. However, the analogue of Proposition \ref{prop:sym_extend}
remains true. The following elementary proof can be easily modified to
apply to symmetric operators as well. 

\begin{proposition}
\label{prop:skew_extend}
Let $\Hil_\mathbb{R}$ be a Hilbert space over $\mathbb{R}$, $A:
\Hil_\mathbb{R} \rightarrow \Hil_\mathbb{R}$ a skew-symmetric operator, and
$A^\mathbb{C}$ the corresponding operator acting on $\Hil_\mathbb{R}
\otimes \mathbb{C}.$ The skew-adjoint extensions of $A$ are in
1-1 correspondence with those skew-adjoint extensions of
$A^\mathbb{C}$ whose domain is invariant under the natural complex
conjugation operator \C. 
\end{proposition}
\textpf{Proof.} Let $B$ be a skew-adjoint extension of $A.$ By
construction, $B^\mathbb{C}$ is skew-symmetric and has a $\C$-invariant
domain. The obvious ``definition chasing'' computation shows that
$B^\mathbb{C}$ is in fact a skew-adjoint extension of $A^{\mathbb{C}}.$

Conversely, let $B^\mathbb{C}$ be a skew-adjoint extension of
$A^\mathbb{C}$ whose domain is invariant under $\C$. First, recall
that $[\C, \left(A^\mathbb{C}\right)^*] = 0$, and notice that, as in
the symmetric case, the skew-symmetry of $A^\mathbb{C}$ implies that
$\Dom\ B^\mathbb{C} \subseteq \Dom\ \left(A^\mathbb{C}\right)^*$. In
other words, $ B^\mathbb{C} =
\left.-\left(A^\mathbb{C}\right)^*\right|_{\Dom
B^{\mathbb{C}}}.$\footnote{The agreement of $B^{\mathbb{C}} $ with
$-\left(A^{\mathbb{C}}\right)^*$ is the only place in this proof where
the assumption of skew-symmetry is needed. For the symmetric case,
Proposition \ref{prop:sym_extend}, it is replaced by the condition
$B^{\mathbb{C}}$ agrees with $\left(A^{\mathbb{C}}\right)^*$. Von
Neumann and Stone's original analysis of real symmetric operators was
based on detailed properties of the Cayley transform, and thus their
results could not be easily extended to skew-symmetric operators.}  This, combined with
the assumed $\C$-invariance of the domain, establish that $[\C,
B^\mathbb{C}] = 0$.  Define \mbox{$B: \Hil_\mathbb{R} \rightarrow
\Hil_\mathbb{R},$} with $\Dom\ B = $ \mbox{$\Dom\ B^\mathbb{C} \cap
\Hil_{\mathbb{R}}$}, by $Bu = B^\mathbb{C}u.$ This operator is well
defined because $[\C, B^\mathbb{C}] =0$, and clearly $B$ extends
$A$. Another straightforward computation establishes that $B$ is
skew-adjoint.
$\blacksquare$

\vspace{1.6ex}
Equivalently, since composition with $i$ does not change the domain,
the skew-adjoint extensions of $A$ are in 1-1 correspondence with the
self-adjoint extensions of $iA^\mathbb{C}$ with \C-invariant
domain. Here, a key a difference between self-adjoint and skew-adjoint
operators arises. Since $iA^{\mathbb{C}}$ anticommutes, rather than
commutes, with $\C$, $\C$ relates the spaces $K_{\pm}$ to themselves
rather than to each other. Thus, $iA^{\mathbb{C}}$ may not have any
self-adjoint extensions. A standard example is the operator $A \psi =
-\frac{d}{dx}\psi$ on $\Dom\ A = \C_0^\infty\left(\mathbb{R}^+\right)
\subseteq L^2\left(\mathbb{R}^{+}, \mathbb{R}\right)$. $A$ is
skew-symmetric, but $iA^\mathbb{C}$ has deficiency indices $n_+ = 0,
n_-=1$ in $L^2\left(\mathbb{R}^+, \mathbb{C}\right)$, so that $A$
possesses no skew-adjoint extensions. 
A key step in later sections will be establishing, in Theorem
\ref{thm:exist}, that skew-adjoint extensions of the time-evolution
operator do indeed exist.

\section{The Prescription}
\label{sec:prescription}

Let $(M,g_{ab})$ be a non-globally-hyperbolic, connected,
stably causal, stationary spacetime. A crucial property of these
spacetimes for this prescription is the following:
\begin{proposition}
If $(M, g_{ab})$ is a connected,
stably causal, stationary spacetime, then there exists a smoothly
embedded spatial hypersurface $\Sigma_0$ which intersects each orbit of the
Killing field exactly once. 
\end{proposition}
\textpf{Proof.} Since $(M, g_{ab})$ is stably causal, it posseses a
smooth global time function \cite{seifert}, i.e., a function $\tau : M
\rightarrow \mathbb{R}$ such that $\nabla^a \tau$ is a past directed
timelike vector field. Consider $\tau=0$, and without loss of generality
take the Killing field $\left(\frac{\del}{\del t}\right)^a$ to be
future directed. As $\nabla^a \tau$ is past directed, it follows that $\tau$
is a strictly increasing function on the orbits of
$\left(\frac{\del}{\del t}\right)^a$, and hence no orbit can intersect
$\tau=0$ more than once. On the other hand, suppose, to the contrary,
that there is some orbit $O$ of $\left(\frac{\del}{\del t}\right)^a$
which does not intersect $\tau=0$. Again without loss of generality, take
$O$ to be entirely to the future of $\tau=0$. Let $B = \dot I^+(O)$, the
boundary of the chronological future of $O$. $B$ is non-empty since
the open set $\tau<0$ does not intersect the open set $I^+(O)$, and $M$
is connected. Let $\mathcal{I}_t$ be the isometry
corresponding to flowing each point of $M$ by Killing parameter
$t$. Since $\left(\frac{\del}{\del t}\right)^a$ is complete,
$\mathcal{I}_t(O) = O \:\forall\: t.$ Furthermore, $\mathcal{I}_t
\left(I^+(O)\right) = I^+\left(\mathcal{I}_t(O)\right) = I^+(O).$ This
means that $\mathcal{I}_t(B) = B \:\forall \:t$, or, equivalently,
that $\left(\frac{\del}{\del t}\right)^a$ is tangent to $B$.  However,
$B$, as the boundary of the chronological future of a set, is
generated by null geodesics, and so cannot have a timelike tangent
vector.  Therefore, each orbit of $\left(\frac{\del}{\del t}\right)^a$
must intersect $\tau=0$ exactly once. $\tau=0$ is also automatically
smoothly embedded by the inverse function theorem, so it may be taken
to be the surface $\Sigma_0$.  $\blacksquare$

\vspace{1.6ex}
\noindent It may appear that the assumption of stable causality is an
overly restrictive causality condition. However, the key property of
$M$ from the standpoint of the prescription is the existence of the
initial surface $\Sigma_0$. Furthermore, any stationary spacetime
containing such a surface is automatically stably causal, as the
Killing parameter may be used to define a global time function on
$M.$ Hence, stable causality is precisely the right 
causality assumption.

Denote by $\gamma_{ab}$ the
Riemannian metric induced by $g_{ab}$ on $\Sigma_0$, $(\Sigma,
\gamma_{ab})$ the abstract manifold so defined, and $(\Sigma_t,
\gamma_{ab})$ the translation of $\Sigma_0$ by Killing parameter $t$
(which need not be related to the time function $\tau$ used to
define $\Sigma_0$).
The lapse function $\alpha$ and shift-vector $\beta^a$ of the Killing
field $\left(\frac{\del}{\del t}\right)^a$ with respect to the future
directed unit normal
$n^a$ of $\Sigma_0$ in $M$ are defined by
\begin{equation*}
\left(\frac{\del}{\del t}\right)^a = \alpha n^a + \beta^a.
\end{equation*}
The classical Hamiltonian for a
Klein-Gordon field with mass $m$ in this foliation is given by
\begin{equation}
\label{eq:hamiltonian}
H((\varphi,\pi)) = \frac{1}{2} \int_\Sigma d\gamma
\alpha\left(\left[1-\frac{\beta_a\beta^a}{\alpha^2}\right]\pi^2
+\gamma^{ab}\left[ D_a\varphi+\frac{\beta_a}{\alpha}\pi\right]
\left[D_b\varphi + \frac{\beta_b}{\alpha}\pi\right] + m^2\varphi^2\right)
\end{equation}
on $C_0^\infty(\Sigma)\oplus C_0^\infty(\Sigma)$, where $\pi :=
n^a\nabla_a \varphi$ is the canonical momentum, $D_a$ is the
Levi-Civita connection of $\gamma_{ab}$, and $d\gamma = \sqrt{\det
\gamma_{\mu\nu}} dx^1 dx^2 dx^3$ is the metric volume form.  Note that
$\beta^a$ is tangent to $\Sigma,$ so it can be pulled back to $\Sigma$
and the above formula makes sense.  This Hamiltonian may be expressed
on $\cipci$ data as $$H((\varphi, \pi)) = \frac{1}{2}\int_\Sigma d\gamma
\left[\begin{array}{cc}\varphi & \pi\end{array}\right] \A
\left[\begin{array}{c}\varphi\\ \pi\end{array}\right],$$ where $\A$ is
the matrix differential 
operator
\begin{equation}
\label{eq:A}
\A = \left[\begin{array}{cc}  -(D^a
    \alpha)D_a + \alpha(m^2 -  D^a D_a)
    & -( D_a\beta^a) - \beta^a D_a \\
     \beta^a D_a & \alpha\end{array}\right].
\end{equation}
Let \HilA\ be the completion of $\cipci$  in the energy norm
\begin{equation}
\label{eq:A-norm}
\|\Phi\|_\A^2 := 2H(\Phi) = \bracket{\Phi}{\A\Phi}_{L^2(d\gamma)\oplus
L^2(d\gamma)},~~~~ \Phi = (\varphi, \pi) \in C_0^\infty(\Sigma)\oplus
C_0^\infty(\Sigma).
\end{equation}
Thus, the squared norm of a field configuration is equal to twice its
classical energy, justifying the name ``energy norm.''

It should be noted that while $\HilA$ is a natural choice of Hilbert
space associated to the system, other choices are possible. That is,
it is possible to define an energy norm which agrees with the above
definition on smooth data of compact support but is defined on a
larger set of functions. In Section \ref{sec:static}, it will be shown
that other choices for $\HilA$ can be used to define different
dynamics, although the definitions of $\A$ and $h$ (below) would then
need to be modified somewhat.

Let $j$ be the standard $2\times 2$ symplectic matrix,
\begin{equation*}
j := \left[\begin{array}{cc}0 & 1\\-1 & 0\end{array}\right],
\end{equation*} and let
\begin{equation}
\label{eq:h}
h := 
-j \A= 
\left[\begin{array}{cc} 
-\beta^a D_a &
-\alpha\\
-(D^a\alpha)D_a + \alpha(m^2 -  D^a D_a) &
 -( D_a\beta^a)  -\beta^a D_a
\end{array}\right].
\end{equation}
Hamilton's equations for the scalar field then take the simple form
$$\frac{\del}{\del t} \Phi(t) = -h \Phi(t).$$  It is easy to check by
partial integration that $h$, defined as a differential operator on
$\cipci \subseteq \HilA$, is a skew-symmetric operator. Formally, the
time-evolution equation may be solved by exponentiating $h$. However, 
this may not be possible unless $h$ is extended to a skew-adjoint
operator $h^{SA}$. Skew-adjoint operators have a spectral
decomposition similar to that of symmetric operators (see, e.g., \cite[Theorem
13.33]{rudin}). This may be used to exponentiate $h^{SA}$ and give
the  solution
\begin{equation}
\label{eq:phi_t}
\Phi_t = e^{-h^{SA} t} \Phi_0. 
\end{equation}
Equation \Ref{eq:phi_t} assigns to each vector in $\HilA$ a
strongly continuous one-parameter family of vectors in $\HilA$ which
represents a formal solution to the Klein-Gordon equation.  However, it
is not  clear \textit{a priori} that (I) this one-parameter family defines
a smooth function on spacetime for smooth initial data of compact
support, or that (II) it solves the Klein-Gordon equation in the pointwise
(spacetime) sense when the initial data are sufficiently
regular. 

In order to make any progress in proving (I) and (II), two technical
assumptions are made. It is possible that these assumption are not
needed, but the methods of proof used Section \ref{sec:properties}
break down. The first assumption restricts consideration to the
massive Klein-Gordon equation:
$$\refstepcounter{condition}
\label{cond:mass} \thecondition~~m^2 > 0. $$ 
 \ref{cond:mass} guarantees that any $\Phi \in \HilA$ is
automatically in $\h{1}\oplus\h{0}$, where $\h{k}$ denotes the $k$th
local Sobolev space (see, e.g., \cite{reed_simonII}). This ensures that all the vectors in $\HilA$ are
functions and not distributions. The second assumption is a
condition on $\alpha$ and $\beta$:
$$
\refstepcounter{condition}\label{cond:beta}
\thecondition~~\alpha - \frac{\beta_a\beta^a}{\alpha} \geq \epsilon > 0.
$$ Notice that since $\alpha$ is manifestly positive, this condition
implies $\alpha \geq \epsilon$ as well as $\left(\frac{\del }{\del
t}\right)^{a}\left(\frac{\del }{\del t}\right)_a = \alpha^2 - \beta^2
> \epsilon^2 > 0.$ However, if $\alpha$ diverges then \ref{cond:beta}
is strictly stronger than the condition that the norm of Killing is
bounded away from zero. Also notice that since it involves $\beta^a$,
\ref{cond:beta} is a condition on the slice $\Sigma_0$ as well as the
Killing field. Roughly speaking, \ref{cond:beta} implies that the
``Lorentz boost'' taking the unit normal $n^a$ to
$\left(\frac{\del}{\del t}\right)^{a}$ is bounded, so that the two
vectors do not ``approach being null with respect to one another.''

A straightforward computation shows that these two conditions
together ensure that $\|\Phi\|_\A \geq c\|\Phi\|_{L^2(d\gamma) \oplus
L^2(d\gamma)}$, where $c$ is some constant depending only on
$\epsilon$ and $m$.  One easy consequence of this
inequality is that the symplectic form on smooth, compactly supported
functions, defined by
\begin{equation}\label{eq:sigma}
\sigma\left(F,G\right) := \bracket{F}{jG}_{L^2(d\gamma) \oplus L^2(d\gamma)},
\end{equation}
is a continuous bilinear form on $\HilA$, i.e., $\exists$ some
constant $c_\sigma$ so that $|\sigma(F,G)| \leq c_\sigma \|F\|_\A
\|G\|_\A.$ This inequality follows from the fact that $j$ is a bounded
operator on $L^2(d\gamma) \oplus L^2(d\gamma)$ combined with the fact
that $\A$-norm bounds $L^2(d\gamma) \oplus L^2(d\gamma)$-norm, and it
means that $\sigma$ extends uniquely to a continuous bilinear form on
all of $\HilA$. This will be critical in in the analysis of Section
\ref{sec:properties}, which shows that (I) is true and that (II) is
true so long as $\Phi_0$ is smooth and $\Phi_0 \in \Dom\ h^{SA}$.
 
To summarize, the prescription is as follows. First, choose any
smoothly embedded spatial hypersurface which intersects each orbit of
the Killing vector field exactly once and satisfies
\ref{cond:beta}. Second, define the Hilbert space $\HilA$ with the
norm given by equations \Ref{eq:A} and \Ref{eq:A-norm}, as well as the
operator $h$ on $\cipci$ by \Ref{eq:h}. Finally, choose an arbitrary
skew-adjoint extension of $h$ and assign the solution \Ref{eq:phi_t}
to each initial datum $\Phi_0 \in \HilA$. Naively, then, two arbitrary
choices---the choice of initial surface and the choice of skew-adjoint
extension---plus a possible third choice, if $\HilA$ may be replaced
by some larger Hilbert space, are allowed in the
prescription. However, as shown in Section \ref{sec:static}, $h$ is
essentially skew-adjoint in static spacetimes. Even if $h$ is not
always essentially skew-adjoint, Theorem \ref{thm:exist} shows that
there is always a preferred skew-adjoint extension. Both these facts
suggest that there is considerably less freedom available than at
first appears. Finally, whether the prescription gives rise to the
same dynamics when inequivalent slicings in the same spacetime are
used remains an open question.

\section{Properties of the Prescription}
\label{sec:properties}
The goal of this section is to prove two theorems. The first assumes
the existence of skew-adjoint extensions of $h$, and proves that any
such extension gives rise to a prescription with reasonable
properties. The proof is more or less a direct adaptation of the
corresponding theorem in \cite{nghsdI}. The second theorem establishes
the existence of skew-adjoint extensions as well as a certain
preferred extension. As noted in Section \ref{sec:operators}, such a theorem was not necessary
in \cite{nghsdI}.

\begin{theorem}
\label{thm:main}
Let $(\Sigma_t, \gamma_{ab})$ be a foliation of $(M,g_{ab})$ which
satisfies \ref{cond:beta}, and consider a minimally coupled
Klein-Gordon equation obeying \ref{cond:mass}. Let $h^{SA}$ be any
skew-adjoint extension of $h$, let \mbox{$\Phi_0 = (\varphi_0, \pi_0)
\in \Dom\ h^{SA}$} be smooth initial data,\footnote{As the proof of
Proposition \ref{prop:agree} makes clear, $\Phi_0 \in C^2(\Sigma)
\oplus C^1(\Sigma)$ would suffice to show $\varphi=\psi$ in
$D(\Sigma_0)$; however, this lower regularity result is not needed in
later sections so for simplicity attention is restricted to smooth
data. In any event, the condition that $\Phi_0 \in \Dom\
\left(h^{SA}\right)^k$ for all $k$ implies that $\Phi_0$ is smooth.}
let $\Phi_t = (\varphi_t, \pi_t)$ be the one-parameter family of
vectors \Ref{eq:phi_t}, and let $\psi$ be the maximal Cauchy evolution
of $\Phi_0$. Define $\varphi : M \rightarrow \mathbb{R}$ by
$\varphi(p,t) = \varphi_t(p)$. Then $\varphi = \psi$ within the domain
of dependence $D(\Sigma_0).$ Further, if $\Phi_0 \in \Dom\
\left(h^{SA}\right)^k\ \forall\: k\in \mathbb{N},$ then $\varphi$ is a
smooth function on $M$ obeying the Klein-Gordon equation. In
particular, smooth data of compact support give rise to smooth
solutions.
\end{theorem}
The proof of this theorem proceeds in three steps. Proposition
\ref{prop:agree} shows that $\Phi_t$ and $\Psi(\cdot, t) :=
(\psi(\cdot, t), n^a\nabla_a \psi(\cdot, t))$ agree almost everywhere
within the domain of dependence. This is a meaningful comparison
because, as noted above, $\Hil_A$ contains only functions. The
continuity of the symplectic form, guaranteed by the assumptions (A) and (B),
is crucial to the proof of this first proposition.  Since
$\Psi_t$ is necessarily smooth, agreement almost everywhere shows
that $\Phi_t$ restricted to the domain of dependence is smooth. Next,
Proposition \ref{prop:smooth} shows that $\Phi_t$ is a smooth function
on each spatial slice. Unlike Proposition \ref{prop:agree}, its proof
does not depend at all on assuming 
\ref{cond:beta}.\footnote{An analysis similar to that preceding Proposition
  \ref{prop:sym_extend} shows that a skew-symmetric operator $A$
  has skew-adjoint extensions if there exists a norm-preserving linear
  involution which anticommutes with $A$. This makes it
  possible to prove that $h$ has skew-adjoint extensions in a
  stationary axisymmetric spacetime with time-reflection symmetry
  without assuming (A) and (B). While, as noted in the
  text, the proof of Proposition \ref{prop:smooth} continues to hold
  in this case, the proof of Proposition \ref{prop:agree}, and hence of
  Theorem \ref{thm:main}, breaks down. It is not known whether this
  breakdown is an artifact of the method of proof or if the result
  fails in this case.}  
Finally, the two results are combined to prove the theorem.
\begin{proposition}
\label{prop:agree}
Under the conditions of Theorem \ref{thm:main}, if $\Psi$ is any
smooth solution of the Klein-Gordon equation with $\Psi|_{\Sigma_0}
\in \Dom\ h^{SA}$, and $\Phi_t :=
e^{-\left(h^{SA}\right)t}\Psi|_{\Sigma_0} ,$ then \mbox{$\Phi_t(p) =
\Psi(p, t)$} a.\ e.\ in $D(\Sigma_0)$ on each slice $\Sigma_t$. Thus, in
particular, they agree  a.\ e.\ throughout $D(\Sigma_0)$.
\end{proposition}
\textbf{Proof.}  Suppose, to the contrary, that $\exists\ t_1$ such
that $\Phi_{t_1}|_{D(\Sigma_0)}$ and $\Psi(\cdot, t_1)|_{D(\Sigma_0)}$
do not agree almost everywhere.  This supposition implies there is
some measurable subset $\Delta \subseteq D(\Sigma_0) \cap
\Sigma_{t_1}$ with positive measure where $\Phi_{t_1}(p) \neq \Psi(p,
t_1).$ Let $S$ be a Cauchy surface for $D(\Sigma_0)$ which coincides
with $\Sigma_1$ on an open set $O$ such that $O \cap \Delta$ has
positive measure.

\begin{figure}[htbp]
\begin{center}
 \includegraphics[height=2.5in]{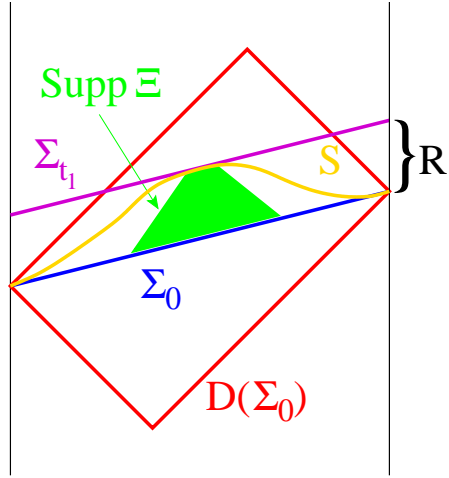}
\end{center}
\caption{An illustration of the proof that $\Xi$ is smooth.  The
  shaded region is $\supp\ \Xi$, which is compactly contained in
  $J^-(S)$. 
}
\label{fig:agree}
\end{figure}

Without loss of generality, take $t_1 > 0$, denote by $J^+$ ($J^-$)
the causal future (past) of a set, and let $\sigma$ be the symplectic
form \Ref{eq:sigma} on $\cipci$ functions.  As noted in Section
\ref{sec:prescription}, assumptions (A) and (B) mean that
$\sigma$ extends uniquely to a continuous bilinear form on all of
$\HilA$, also denoted $\sigma$.  Choose $\tilde \Xi \in C_0^\infty(O)
\oplus C_0^\infty(O)$ such that $\sigma(\tilde \Xi, \Phi_{t_1} -
\Psi(\cdot, t_1)) \neq 0.$ Such a $\tilde \Xi$ exists because $\sigma$
is a non-degenerate bilinear form. Now extend $\tilde \Xi$ to $\Xi$, a
smooth solution of the Klein-Gordon equation defined in the entire
region \mbox{$R := J^+(\Sigma_0) \cap J^-(\Sigma_{t_1})$}, as
follows. Within $D(\Sigma_0)\cap R$, evolve $\tilde \Xi$ by the
ordinary Cauchy evolution; in the remainder of $R$, set it to
vanish. Note that, by construction, there is a compact set $\tilde K$
with \mbox{$\supp ~\tilde \Xi\subseteq \tilde K \subseteq
\mathrm{Int}~\left(S\cap\Sigma_{t_1}\right)$}, and recall that both $S$
and $\Sigma_0$ are Cauchy surfaces for $D(\Sigma_0)$. From this it
follows that there is a compact set $K$ with $\supp
~\Xi|_{D(\Sigma_0)\cap R} \subseteq K \subseteq
\mathrm{Int}~J^-(\Sigma_{t_1}\cap S) \cap J^+(\Sigma_0)\subseteq
\mathrm{Int}~D(\Sigma_0)\cap R,$ which shows that $\Xi$ is
indeed a smooth solution. This situation is illustrated in Figure
\ref{fig:agree}.

Consider now 
\begin{equation}
\label{eq:sigma_t}
\sigma(t) :=\sigma\left(\Xi(\cdot, t), \Phi_{t} -
\Psi(\cdot, t)\right) = \sigma\left(\Xi(\cdot, t), \Phi_{t} \right) -
\sigma\left(\Xi(\cdot, t), \Psi(\cdot, t)\right) 
\end{equation}
By construction, $\sigma(t_1) \neq 0$, whereas $\sigma(0) = 0$ because
$\Phi_0 = \Psi(\cdot, 0)$. A contradiction will now be established by
showing that the derivative of $\sigma$ is zero. Since $\Xi$ and
$\Psi$ are both smooth solutions of the Klein-Gordon in $D(\Sigma_0),$
one of which has compact support, the derivative of the second term of
$\sigma(t)$ automatically vanishes. On the other hand,
$\Xi(\cdot, t) \in \HilA$ for each $t \in [0, t_1]$ because it is
smooth and of compact support. Further, this one parameter family of
vectors is also strongly differentiable in the Hilbert space sense,
with derivative $-h\Xi(\cdot, t)$ \cite{kay}. By the generalized
Stone's Theorem \cite[Theorem 13.35]{rudin}, $\Phi_0 \in \Dom\ h^{SA}
\Leftrightarrow \Phi_t \in \Dom\ h^{SA}$, 
$\Phi_t$ is strongly differentiable, and $\Phi_t' =
-h^{SA}\Phi_t$. Hence the derivative of the first term of $\sigma(t)$ is
$\sigma(-h\Xi(\cdot, t), \Phi_t) + \sigma(\Xi(\cdot, t), -h^{SA}
\Phi_t)$. Let $\{F_n\}\subseteq \cipci$, be a sequence which converges
to $\Phi_t$ in $\HilA$ such that $-h^{SA}F_n \rightarrow
-h^{SA}\Phi_t$. Such a sequence exists because $h^{SA}$ is a closed
operator. Since $\A$ is symmetric on smooth
data of compact support, and $\A$-norm bounds $L^2(d\gamma)\oplus
L^2(d\gamma)$-norm,   
it follows that
\begin{eqnarray}\nonumber
\sigma(-h\Xi(\cdot, t), \Phi_t) + \sigma(\Xi(\cdot, t), -h^{SA}
\Phi_t)& =& \lim_{n\rightarrow \infty} \sigma(-h\Xi(\cdot, t), F_n) +
\sigma(\Xi(\cdot, t), -h^{SA} F_n)\\\nonumber&=&
\lim_{n\rightarrow \infty} \bracket{-h\Xi(\cdot, t)}{j
  F_n}_{L^2\oplus L^2} +
\bracket{\Xi(\cdot, t)}{j (-h) F_n}_{L^2\oplus L^2}
\\\nonumber&=&  
\lim_{n\rightarrow \infty} \bracket{j\A\Xi(\cdot, t)}{j F_n}_{L^2\oplus  L^2}  
+ \bracket{\Xi(\cdot, t)}{j^2 \A F_n}_{L^2\oplus L^2}
\\\nonumber &=&
\lim_{n\rightarrow \infty} \bracket{\A\Xi(\cdot, t)}{ F_n}_{L^2\oplus  L^2}
 -
\bracket{\Xi(\cdot, t)}{\A F_n}_{L^2\oplus L^2}\\\nonumber
&=&0,
\end{eqnarray}
where $L^2$ denotes $L^2(d\gamma).$ This establishes the
contradiction, so $\Phi_t$ and $\Psi(\cdot,t)$ agree a.\ e.\ in
$D(\Sigma_0)$.~$\blacksquare$

\vspace{1.6ex}
While the proof of the next proposition is conceptually similar to the
corresponding result in \cite{nghsdI},  it differs significantly in
detail. The use of the Hilbert space $L^2\left(\alpha^{-1}d\gamma\right)$ in
\cite{nghsdI} allowed a natural identification of the solution
\Ref{eq:static_sol} with a distribution in \h{0}, and then the strong
ellipticity of the operator $A$ was used to show the regularity of
$\phi_t$. The key to the proof below is the identification of the
distribution $X$ (defined below) as a distribution in the negative
order Sobolev space $\h{-1} \oplus
\h{0}$. Further, rather than utilizing the ellipticity of the
operator $\A h$ directly, its detailed form is used to identify an
elliptic operator acting on a piece with sufficient \textit{a priori}
regularity to allow iteration. 

\begin{proposition}
\label{prop:smooth}
Let $\Phi_t =:
\left(\varphi_t, \pi_t\right)$ be the one-parameter family
of vectors \Ref{eq:phi_t} corresponding to $\Phi_0$, and assume
\ref{cond:mass} holds. If $\Phi_0 \in
\Dom\ \left(h^{SA}\right)^k\ \forall\: k \in \mathbb{N}$, then
$\varphi(\cdot, t) := \varphi_t$ and $\pi(\cdot, t) := \pi_t$ are
smooth functions on $\Sigma_t$ for each fixed $t$.
\end{proposition}
\textpf{Proof.} Suppose $\Phi_0 \in \Dom\ h^{SA}$. By Stone's Theorem
$\Phi_t \in \Dom\ h^{SA}$, which implies that $h^{SA} \Phi_t \in
\h{1} \oplus \h{0}.$ Consider the distribution $$X(F) :=
-\bracket{h F}{\Phi_t}_\A.$$ By the
skew-symmetry of $h^{SA}$ in $\HilA$, $X(F) = \int d\gamma F^T \A h
\Phi_t$, where $T$ denotes matrix transpose and $\A$ and $h$ are now
viewed as a differential operator 
on distributions.  Hence, $X$, $\varphi_t$, and $\pi_t$ satisfy the
distributional differential equation
\begin{equation}
\label{eq:smootha}
X = \A h \Phi_t = \left[\begin{array}{c}
\left(\alpha^2 \gamma^{ab} + \beta^a\beta^{b}\right)D_aD_b \pi_t 
+ (D_a \beta^a)D^bD_b\varphi_t + \textrm{lower-order terms} \\
\left(\alpha^2 \gamma^{ab} + \beta^a\beta^{b}\right)D_aD_b \varphi_t +
\textrm{lower-order terms} \\ \end{array}\right].
\end{equation}
Since $\varphi_t \in \h{1}$ \textit{a priori}, the second term in the
top component of the right hand side of \Ref{eq:smootha} is in
$\h{-1}.$ The lower-order terms on the right hand side are also
contained in $\h{-1}$. On the other hand, the top left component of
$\A$ contains two derivatives and the top right component only one,
so $h^{SA}\Phi_t \in \h{1} \oplus \h{0}$ implies that the top
component of $X$ is in $\h{-1}.$ Hence,
\begin{equation}
\label{smoothb}
E\pi_t := (\alpha^2\gamma^{ab}+\beta^a\beta^b)D_a D_b \pi_t \in
\h{-1}.
\end{equation}
As $E$ is clearly an elliptic second-order differential operator,
$\Ref{smoothb} \Rightarrow \pi_t \in \h{1}$. With both $\pi_t$ and
$\varphi_t$ established as being in $\h{1}$, the lower-order terms in the
bottom component of the right hand side of Equation \Ref{eq:smootha} are
at least \h{0}. Since the bottom left component of
$\A$ contains one derivative and the bottom right component none, 
 $h^{SA}\Phi_t \in \h{1} \oplus \h{0}$ implies that the bottom
component of $X$ is in $\h{0},$
and hence $E \varphi_t \in \h{0} \Rightarrow \varphi_t \in \h{2}.$ Suppose now
that $\Phi_0 \in \Dom\ \left(h^{SA}\right)^k$. It is clear that the
above argument could be repeated, peeling one $h^{SA}$ at a
time. Hence $\varphi_t \in \h{k+1}$ and $\pi_t \in \h{k}$. As this is
true $\forall\: k$ by assumption, both $\varphi_t$ and $\pi_t$ are
smooth.~$\blacksquare$

\vspace{1.6ex} \textpf{Proof of Theorem \ref{thm:main}.} It remains
only to show that $\varphi$ is smooth solution of the Klein-Gordon
equation in $M$.\footnote{In the static case, it is necessary to argue
that $\frac{d\phi}{dt}$ is smooth on each spatial slice before
spacetime smoothness of the $\phi$ can be established
\cite{nghsdI}. In the present case, this is not necessary because
$\Phi_0$ by itself completely determines the solution $\Phi_t$.} Let
$p \in M$, and let $\Sigma_{t_p}$ be the slice passing through
$p$. Let $\tilde \Phi_t = e^{h^{SA}(t-t_p)}\Phi_{t_p}$, and let
$\tilde \Psi$ be the maximal Cauchy evolution of $\Phi_{t_p}$ in
$D(\Sigma_{t_p})$. Again by Proposition \ref{prop:agree}, $\tilde
\Phi_t = \tilde \Psi(\cdot ,t)$ in $D(\Sigma_{t_p})$, which certainly
includes some neighborhood of $p$. The definition of $\tilde \Psi$
therefore shows that $\varphi$ is smooth and obeys the Klein-Gordon
equation in some neighborhood of $p$. As $p$ is arbitrary, the result
follows.~$\blacksquare$

\begin{theorem}\label{thm:exist}
Under the conditions of Theorem \ref{thm:main}, $h$ possesses a
skew-adjoint extension which has a bounded inverse and 
conserves  the natural symplectic form $\sigma$.
\end{theorem}

\textpf{Proof.} Because the symplectic form $\sigma$ is continuous
with respect to $\A$-norm,
the Riesz lemma shows $\exists$ a bounded skew-adjoint operator
$T:\HilA \rightarrow \HilA$ such that $\sigma(\Phi, \Psi) =
\bracket{\Phi}{T\Psi}_\A \:\forall~\Phi, \Psi \in \HilA$. Further, 
$\forall\: F,G \in \cipci$,
$$\bracket{F}{ThG}_\A = \sigma(F, hG) = \bracket{F}{jhG}_{L^2(d\gamma)
  \oplus L^2(d\gamma)} = \bracket{F}{-j^2 AG}_{L^2(d\gamma)\oplus
  L^2(d\gamma)}= \bracket{F}{G}_\A,$$ or that $Th = 1$ on \cipci. This
  calculation also shows that $\mathrm{Ran}~T$ is dense. Because $T$
  is skew-adjoint, this suffices to show that $T^{-1}$ exists and is
  skew-adjoint \cite[Theorem 13.11(b)]{rudin}. But $T^{-1}|_{\cipci} =
  h$, so that $T^{-1}$ is a skew-adjoint extension of $h.$ $T$ is by
  construction bounded, and clearly the symplectic form is conserved
  by the time evolution associated to $T^{-1}$ because
  $[T,T^{-1}]=0$.~$\blacksquare$

\section{Static Spacetimes}
\label{sec:static}
It is interesting to compare the first-order formalism in the static
slicing with the previously studied second-order formalism
\cite{nghsdI,nghsdII, horowitz_marolf, naked_singularities}. From this
analysis will come the result that the definition of the Hilbert space
determines the dynamics in these spacetimes. This suggests that the
choice of Hilbert space in a general stationary spacetime plays at
least as large a role, if not a larger role, in determining the
dynamics as does the choice of skew-adjoint extension $h$.

In the static slicing, $\beta^{a} = 0$ so $\A$ and $h$ take the form
\begin{equation}
\label{eq:staticah}
\A = \left[\begin{array}{cc} \alpha^{-1}A & 0 \\ 0 &
    \alpha\end{array}\right]~~~~~~~
h = \left[\begin{array}{cc}  0 & -\alpha \\ \alpha^{-1}A & 0\end{array}\right],
\end{equation}
where $A$ is the operator \Ref{eq:static_eq}. Thus, $\HilA$ decomposes
as a direct sum $\HilA = \X \oplus \Y$, where $\X = H^1_0(\alpha
d\gamma)$ and $\Y = L^2(\alpha d\gamma)$. Another space which plays a
role in the analysis is $\Z:= L^2(\alpha^{-1}d\gamma)$; this is the
Hilbert space in which $A$ is a symmetric operator. Note that $\pi
\rightarrow \alpha \pi$ maps $\Y$ bijectively onto $\Z$. This is
merely a restatement of the fact the canonical momentum is $\pi =
n^a\nabla_a \varphi$, whereas the time derivative of $\varphi$ is
$\frac{\del \varphi}{\del t} = \alpha n^a\nabla_a \varphi =
\alpha\pi$.

The relationship between $\X$ and $\Z$ is slightly more
complicated. $\X$ is the completion of $C_0^\infty$ in the norm
$\|f\|_\X^2 = \bracket{f}{A f}_\Z$. Clearly $\|f\|_\X^2 \leq
\|f\|_{A^{\frac{1}{2}}} := \bracket{f}{A f}_\Z + \|f\|_\Z^2$. However,
if \ref{cond:beta} holds so that $\alpha \geq \epsilon > 0$, then it
is also true $(1+\epsilon^{-2})^{-1}\|f\|_{A^{\frac{1}{2}}} \leq
\|f\|_\X.$ Hence $\X$-norm and $A^{\frac{1}{2}}$-norm are
equivalent. The completion of $C_0^\infty$ in $A^{\frac{1}{2}}$-norm
is $q_F$, the form-domain of the Friedrichs extension of $A$
\cite{reed_simonII}. $q_F$ is also equal to $\Dom\ A_F^{\frac{1}{2}}$,
where $A_F$ denotes the Friedrichs extension of $A$
\cite{reed_simonI}.
These relationships between \X, \Y, and \Z\ are the key understanding
the dynamics in static spacetimes.

\begin{theorem}
\label{thm:esa}
Let $(M,g_{ab})$ be a static spacetime obeying \ref{cond:beta} in the
static slicing. If \ref{cond:mass} holds, then $h$ is essentially
skew-adjoint.
\end{theorem}
\textpf{Proof.}
The deficiency indices of $ih^{\mathbb{C}}$ are computed directly and
shown to be zero. Suppose, to the contrary, that $0 \neq \Phi = (\varphi,
\pi) \in K_\pm$. This means that $\forall\: F = (f, g) \in \cipci,$
$\Phi$ satisfies $\int d\gamma \Phi^T \A (ih^{\mathbb{C}} \pm i) F =
0$. The preceding discussion shows that the map $(\varphi,
\pi)\rightarrow (\varphi, \alpha \pi)$ injects $\HilA \hookrightarrow \Z
\oplus \Z$, so that the index equations can be written as
\begin{eqnarray}
\label{eq:index_first}
-\bracket{\varphi}{Ag}_\Z \mp \bracket{\alpha\pi}{g}_\Z & = & 0\\
\label{eq:index_second}
\bracket{\alpha\pi}{Af}_\Z \mp \bracket{\varphi}{Af}_\Z & = & 0.
\end{eqnarray}
Equation \Ref{eq:index_first} is nothing but the statement $\varphi \in
\Dom\ A^*$ (where the adjoint is taken in \Z) and $A^*\varphi = \mp \alpha
\pi$. Equation \Ref{eq:index_second} then becomes $A^*(\alpha \pi) =
\pm A^* \varphi.$ Clearly, one solution is $\alpha \pi = \varphi$, or $A^*
\varphi = -\varphi$. Suppose that $\alpha \pi = \varphi + \delta$ were a
second solution. This would mean that $A^*\delta = 0$, or that
$\bracket{F}{(\delta,0)}_\A = 0 \:\forall\: F \in \cipci.$ As $\cipci$
is dense in $\HilA$, it follows that $(\delta, 0) = 0$ in $\HilA.$
Thus $A^*\varphi=-\varphi,~\alpha \pi = \varphi$ is the unique solution to
\Ref{eq:index_first}, \Ref{eq:index_second}. Now, let $F_n :=
\{(f_n, g_n)\} \subseteq \cipci$ be a sequence which converges
to $\Phi$ in $\HilA$. Since $\|F_n\|_\A \geq
\|F_n\|_{\Z\oplus\Z}$, it follows that $f_n \rightarrow
\varphi$ in $\Z$. But $\X = q_F$, which means that $\|(f_n, 0)\|_\A
\rightarrow \bracket{\varphi}{A\varphi}_\Z = -\|\varphi\|_\Z^2$, which
is clearly impossible. Hence $K_\pm = \{0\}$ and $h$ is essentially
skew-adjoint.~$\blacksquare$

\vspace{1.6ex} In \cite{nghsdII} it was shown that any reasonable
prescription agrees with the second-order formalism on $C_0^\infty$
data for some choice of positive, self-adjoint extension $A_E$ of
$A$. The following theorem shows, without invoking the result of
\cite{nghsdII}, that the first-order formalism agrees with the second
order formalism when the Friedrichs extension is chosen. It actually
shows a stronger result, namely that the prescriptions agree for data
in $\HilA.$

\begin{theorem}
\label{thm:friedrichs}
Let $(M,g_{ab})$ be a static spacetime obeying \ref{cond:beta}, and
suppose that \ref{cond:mass} holds. 
In the static slicing define $U^{(1)} : \HilA \rightarrow \HilA$ by
$U^{(1)} = e^{-\bar h t}$, where $\bar h$ 
denotes the operator closure of $h$ in $\HilA$, and let $U^{(2)} : \Z
\oplus \Y \rightarrow \Z \oplus \Y$ be given by
$$U^{(2)}(t) = 
\left[\begin{array}{cc}
\cos\left(A_F^{\frac{1}{2}}t\right) &
A_F^{-\frac{1}{2}}\sin\left(A_F^{\frac{1}{2}}t\right)\alpha \\ 
-\alpha^{-1} A_F^{\frac{1}{2}}\sin\left(A_F^{\frac{1}{2}}t\right)&
\alpha^{-1}\cos\left(A_F^{\frac{1}{2}}t\right)\alpha \\ 
\end{array}\right],$$
 where $A_F$ is  the
Friedrichs extension of $A$ in \Z. Then $U^{(2)}(-t) U^{(1)}(t)$ is a
well-defined map $\HilA \rightarrow \HilA$, and further
\begin{equation}
\label{eq:upsilon_t}
U^{(2)}(-t) U^{(1)}(t) = \mathbb{I}_{2\times 2}.
\end{equation}
\end{theorem}
\textpf{Proof.} Notice that the operators in \Ref{eq:upsilon_t}
correspond to evolving the initial state by a time 
$t$ in the first-order formalism, then evolving time backward by $t$
in the second-order formalism. Thus, if $U^{(2)}(-t)U^{(1)}(t)$ maps
$\HilA$ to itself and further $U^{(2)}(-t)U^{(1)}(t) =
\mathbb{I}_{2 \times 2}$, then these two evolutions invert each other,
and hence the first-order and second-order formalism agree. 

Let $(\varphi_0, \pi_0) \in \HilA$ and define $$\Upsilon_t :=
U^{(2)}(-t)U^{(1)}(t) (\varphi_0, \pi_0).$$ It will at first be
assumed that $(\varphi_0, \pi_0) \in \Dom\ \bar h$. This assumption
will be used to show that $\Upsilon_t$ differentiable in $\Z \oplus
\Y$ and, in fact, constant.  It will then be shown that $\Upsilon_t$
is constant for arbitrary $(\varphi_0, \pi_0) \in \HilA$, which
establishes the result \Ref{eq:upsilon_t}.

It is first necessary to show that $U^{(2)}(-t) U^{(1)}(t)$ maps
$\HilA \rightarrow \HilA$ so that $\Upsilon_t$ is
well-defined. Because Theorem \ref{thm:esa} shows that $h$ is
essentially skew-adjoint, its closure $\bar h$ is skew-adjoint. This
suffices to show that $U^{(1)}(\varphi_0, \pi_0)$ is a well-defined
vector in $\HilA$. As noted above, $\X = q_F = \Dom\
A_F^{\frac{1}{2}}$. Hence the $\X$ component of $\Upsilon_t$ is
well-defined because (i) $\cos(A_F^\frac{1}{2}t)$ is a bounded
operator (on \Z) which leaves $q_F$ invariant, and (ii)
$A^{-\frac{1}{2}}_F\sin(A_F^\frac{1}{2}t)$ is a bounded operator on
$\Z$ with range contained in $\Dom\ A_F^{\frac{1}{2}}$, and $\alpha$
maps $\Y$ to $\Z$.  On the other hand, the $\Y$ component of
$\Upsilon_t$ is well-defined because (iii) $\psi \in \X
\Leftrightarrow \alpha^{-1}A_F^{\frac{1}{2}}\sin(A_F^\frac{1}{2}t)
\psi \in \Y$, and (iv) $\cos(A_F^\frac{1}{2}t)$ is a bounded operator
on $\Z$, combined with the fact that $\alpha$ maps $\Y$ to $\Z$
bijectively. Notice that (iii) and (iv) would continue to hold if
$A_F$ were replaced by $A_E$ in the definition of
$U^{(2)}(t)$. However, (i) would fail for an arbitrary
extension. Further, (ii) necessarily fails for any other extension
$A_E$ because $q_F$ is the closure of the quadratic form associated to $A$
\cite{reed_simonII}. This implies that $\Dom\ A_E^{\frac{1}{2}}
\supsetneq \Dom\ A_F^{\frac{1}{2}}$ \cite{reed_simonI}, and and hence
$A_{E}^{-\frac{1}{2}}\sin(A_E^{\frac{1}{2}}\alpha)$ maps $\Y$ outside
of $\X$.
Therefore this theorem identifies $A_F$ as the unique self-adjoint
extension whose dynamics agree with the first-order formalism.

Suppose now that $(\varphi_0,\pi_0) \in \Dom\ \bar h,$ which by definition is
the closure of $\cipci$ in the norm:
\begin{eqnarray}
\nonumber
\|(f, g)\|_{\bar h}^2& :=& \|(f, g)\|_\A^2 + \|h(f,
g)\|_\A^2\\
\label{eq:h_norm}
& = &  
\bracket{f}{Af}_\Z + \bracket{f}{A^2f}_\Z +
\|\alpha g\|_\Z^2 + \bracket{\alpha g}{A\alpha g}_\Z.
\end{eqnarray}
On the one hand, $\frac{\epsilon^2}{m^2}\bracket{f}{f} \leq
\bracket{f}{Af}$. On the other hand, the Cauchy-Schwartz inequality
and the fact that $A$ is symmetric imply that 
$$\bracket{f}{Af}_\Z \leq \left\|f\right\|_\Z \left\|Af\right\|_\Z
\leq \frac{1}{2}\left(\bracket{f}{f}_\Z + \bracket{Af}{Af}_\Z\right)
=\frac{1}{2}\left(\bracket{f}{f}_\Z + \bracket{f}{A^2f}_\Z\right).$$
Thus, the first two terms in \Ref{eq:h_norm} are norm equivalent to 
$\bracket{f}{f}_\Z + \bracket{f}{A^2f}_\Z,$ which means that $\varphi_0
\in \Dom\ \bar A \subseteq \Dom\ A_F$. The last two terms in
\Ref{eq:h_norm} are operator-closure norm for $A_F^{\frac{1}{2}}$, so
$\alpha \pi_0 \in \Dom\ A_F^{\frac{1}{2}}.$

Let $\Psi = (\psi, \theta)$ be such that $\psi  \in \Dom\ A_F$ and $\theta \in \Y.$ Arguments similar to those
which prove Stone's Theorem as well as the fact that $\alpha$ maps $\Y$
to $\Z$ bijectively show that $U^{(2)}(t)\Psi$ is strongly differentiable in $\Z \oplus \Y$ with
derivative 
$$D^{(2)} U^{(2)} \Psi := \left[\begin{array}{cc} 0 & \alpha \\ -A_F & 0
  \end{array}\right]U^{(2)} \Psi.$$
By the generalized Stone's theorem \cite[Theorem 13.35]{rudin}, the
strong derivative of $U^{(1)}(t) (\varphi_0, \pi_0)$ exists in $\HilA$
and is given by
$$D^{(1)} U^{(1)}(t) (\varphi_0, \pi_0) := -\bar h(\varphi_t, \pi_t) =
(-\alpha\pi_t, \bar A \varphi_t).$$ The last equality follows because
action of $h$ on $f$ agrees with the action of $A$ on $f$ in a core
domain $C_0^\infty$. However, $\HilA$-norm bounds $\Z \oplus \Y$-norm, so
any one parameter family of vectors which is norm differentiable in
$\HilA$ is also norm differentiable in $\Z \oplus \Y$. Hence
$U^{(1)}(t) (\varphi_0, \pi_0)$ is strongly differentiable in $\Z
\oplus \Y.$   The generalized Stone's theorem further shows
that $U^{(1)}(t) (\varphi_0, \pi_0) \in \Dom\ \bar h$ and hence that
$U^{(1)}(t) (\varphi_0, \pi_0) \in \Dom\ \bar A \oplus \Y$ for all $t \in
\mathbb{R}$. Thus, the Leibnitz rule may be used, yielding
\begin{equation}\nonumber
\Upsilon_t' = U^{(2)}(-t) \left[D^{(1)} - D^{(2)}\right] U^{(1)}(t) 
\left(\begin{array}{c}\varphi_0\\\pi_0\end{array}\right), 
\end{equation}
Again because the $\X$ component of $\Dom\ \bar h$ is $\Dom\ \bar A
\subseteq \Dom\ A_F$,
it follows that $D^{(2)}|_{\Dom\ \bar h} = D^{(1)}$, or $\Upsilon'_t = 0.$
Hence $\Upsilon_t$ is a constant equal to $\Upsilon_0 = (\varphi_0,
\pi_0).$ 

It remains only to show that $\Upsilon_t = \Upsilon_0$ for arbitrary
$(\varphi_0, \pi_0) \in \Hil_A.$ However, $U^{(1)}(t)$ is a unitary
transformation by the spectral theorem.  $U^{(2)}(-t)$ is a bounded
operator on $\HilA$, as can be shown by a straightforward computation
utilizing facts (i)-(iv). Since it inverts the unitary operator
$U^{(1)}(t)$ on the dense domain $\Dom\ \bar h$, $U^{(2)}(-t)|_\HilA$
is also unitary and $\left(U^{(2)}(-t)|_\HilA\right)^{-1} =
U^{(1)}(t)$.  Thus, $U^{(2)}(-t)U^{(1)}(t) = \mathbb{I}_{2 \times 2}$
on $\HilA$.~$\blacksquare$

\vspace{1.6ex} Theorem \ref{thm:esa} suggests that the first-order
prescription is unique. However, it is known \cite{horowitz_marolf,
naked_singularities} that $A$ is not essentially self-adjoint in \Z---and
hence the choice of dynamics is not unique---in a variety of static
spacetimes. This raises the question of why the first-order formalism
appears unique, but the proof of Theorem \ref{thm:friedrichs} has
already provided the answer. It appears unique because $\HilA$ was
defined as the completion of $\cipci$ in $\A$ norm. To produce other
dynamics, define $\tilde \Hil_\A := q_E \oplus \Y$, where $q_E$ is the
form domain of any other positive, self-adjoint extension $A_E$ of $A$
in $\Z$. Since $\cipci$ is no longer dense in $\HilA$, one cannot
define $h$ on this domain and take its closure. However, $\tilde h$
may be defined directly in terms of the spectral resolution of
$A_E$. The obvious modification of the proof of Theorem
\ref{thm:friedrichs} then shows that the dynamics defined by $\tilde
h$ agrees with the dynamics defined by $A_E$ on data contained in $q_E
\oplus \Y$. Thus, in the first-order formalism, it is the choice of
Hilbert space that encodes the ``boundary conditions at the
singularity'', as opposed to the second-order formalism where they are
encoded in the choice of self-adjoint extension of $A$.

It should be emphasized, as noted above, that care must be taken to
ensure that an operator defined on an energy-type Hilbert space is in
fact densely defined. It appears that this point was overlooked in the
analysis of \cite{naked_singularities}, leading to a mistaken
conclusion that the operator $A$ of that reference is not essentially
self-adjoint in certain spherically symmetric
spacetimes. In \cite{naked_singularities},  the operator
\Ref{eq:static_eq}, initially defined on $C_0^\infty(\Sigma)$, was
viewed as an operator in $H^1$. However, if
$C_0^\infty$ is not a dense subspace of $H^1$, then $A$ is not
well-defined. On the other hand, if $C_0^\infty$ is dense, then the deficiency
subspaces of $A$ are trivial because the solutions to \mbox{$(A^* \pm
i)\phi = 0$}, while they are square integrable, are not
well-approximated by $C_0^\infty$ functions. This correction actually
strengthens the conclusion of that work, in that it means that all the
spacetimes considered there are ``wave regular,'' that is, possessing
an essentially self-adjoint $A$. This suggests that the essential
skew-adjointness of $h$ in static spacetimes may be a general property of
using Sobolev (energy) Hilbert spaces and not of the first-order
formalism \textit{per se}.

\section{Field Quantization}
\label{sec:quantum}
There is a general prescription for defining a quantum vacuum
associated to any give notion of energy which was originally proposed by
Ashtekar and Magnon \cite{Ashtekar:zn}, further developed by Kay
\cite{kay}, and generalized by Chmielowski \cite{chmielowski}. For a
globally hyperbolic spacetime, the procedure is as follows. Let
$\Sigma$ be a Cauchy surface for the globally hyperbolic spacetime
$(M,g_{ab})$ . Denote by $(S, \sigma)$ the space of classical
solutions, which may be identified with the initial data $\cipci$,
together with the symplectic form, which may be done as the initial
value problem is well-posed in these spacestimes. Let 
$\mu$ be any inner product on $S$ which satisfies the inequality
\begin{equation}
\label{eq:quasifree}
\left|\sigma(\Phi, \Psi)\right| \leq 2 \sqrt{\mu(\Psi, \Psi)\mu(\Phi,\Phi)},
\end{equation}
and let $\Hil_\mu$ denote the completion of $S$ in $\mu$-norm.  Kay
and Wald \cite{kay_wald} showed that $\mu$ defines a state in the
class of quasifree states, which includes both vacuum and thermal
states. ($\frac{1}{2}\mu$ is essentially the real part of the two-point
function of the state. See \cite{kay_wald} or \cite{chmielowski} for
the details.) As $\sigma$ is skew-symmetric and bounded on $\Hil_\mu$,
there is a unique, bounded, skew-adjoint operator $T_\mu$ on
$\Hil_\mu$ such that
\begin{equation}
\label{eq:tmu}
\sigma(\Phi, \Psi) = 2 \mu (\Phi, T_\mu \Psi) \:\forall\: \Phi, \Psi
\in S.
\end{equation}
 Define a new inner product on $S$ by $\tilde \mu (\Phi, \Psi) = \mu
(\Phi, |T_\mu| \Psi),$ where $|T_\mu|$ denotes the ``absolute value''
of $T_\mu$ (see, e.g., \cite{reed_simonI}). The state defined by
$\tilde \mu$ is pure in addition to being quasifree, and hence is a
vacuum state \cite{chmielowski}. Chmielowski
further showed that if $\mu$ is the Hamiltonian \Ref{eq:hamiltonian}
(rescaled so that it saturates the inequality \Ref{eq:quasifree}),
then the vacuum state so defined is the well-known
``frequency-splitting'' ground state associated to a timelike Killing field.

The assumption of global hyperbolicity was used subtly in the above
argument, allowing the identification of $\cipci$ with the space of
solutions. This meant that the symplectic form defined via the
integral \Ref{eq:sigma} was well-defined on the space of solutions $S$
because it is non-degenerate and conserved for solutions of the wave
equation. Similarly, the norm $\mu$ on the space of solutions could be
defined via an integral on some spatial slice because there was a
natural identification of solutions with initial data on that
slice. In the non-globally-hyperbolic case, the situtation is much
more complicated. In the absence of any prescription for selecting
solutions, a given initial datum might be identified with many
different solutions of the equation. Hence, the Hamiltonian
\Ref{eq:hamiltonian} would fail to define an inner product because
there would be non-zero solutions which have zero norm. Worse yet, the
sympletic form \Ref{eq:sigma} would fail to be non-degenerate. Thus,
the quantum vacuum cannot be defined because the space of all
solutions to the wave equation is ``too large.''  By analogy with the
globally hyperbolic space, one might try to define a smaller space
$\tilde S$ by requiring that the solutions be compactly supported on
all spatial slices.  However, the time evolution carries $\cipci$ data
to non-compactly supported data, in general, and thus $\cipci$ is
``too small'' as the initial data set.

However, the prescription for defining dynamics presented above is
automatically defined not on just $\cipci$ but on the whole Hilbert
space $\HilA$. Further, $\sigma$ extends to a bounded bilinear form on
$\HilA$, so that $\HilA$ is a ``small enough'' space to have a
well-defined symplectic structure. If one uses the extension $h^{SA} =
T^{-1}$ (where $T$ is as in the proof of Theorem \ref{thm:exist}),
$\sigma$ is conserved for all solutions with initial data in
$\Dom~h^{SA}$, so that space can be identified with a space of
solutions. Since the prescription automatically conserves $\A$-norm,
if an inner product $\mu$ is defined by $\mu(\Phi, \Phi) =
\bracket{\Phi}{\Phi}_\A$ (plus appropraite rescaling to satisfy
\Ref{eq:quasifree}), then $\tilde \mu = \mu$.  This may be seen
explicitly from the proof of Theorem \ref{thm:exist}, which shows that
$\left|T_\mu\right|= \left|T\right|$. The fact that $h^{SA} = T^{-1}$
has a bounded inverse has an important physical
consequence: the real part of the two-point function in the
quantum vacuum associated to $T^{-1}$ is given by
$$\frac{1}{2}\tilde \mu_{T^{-1}}(\Phi, \Psi) =
\frac{1}{2}\bracket{\Phi}{\left|T\right|\Psi}_\A.$$ However, if $T$
were not bounded, then the smeared two-point function would
diverge.  Hence, $|T|^{-1}$ has a ``mass gap'' and thus avoids ``infrared
divergences'' in its quantum vacuum.

The foregoing analysis, along with the results of
\cite{horowitz_marolf}, \cite{naked_singularities} that in certain
static spacetimes the operator $A$ is not essentially self-adjoint,
shows there are many different ``energy splitting'' ground states
possible in a non-globally-hyperbolic spacetime. In these
static spacetimes, $T$ is a fixed, bounded function of
$A_E^{-\frac{1}{2}}$, where $A_E$ is the self-adjoint extension used
to define the Hilbert space $\tilde \Hil_\A$. Since different
extensions have different spectral decompositions (on $\Z$) and
$A_E^{-\frac{1}{2}}$ is a bounded operator, it follows that $T$ is
different for different choices of $\tilde \Hil_\A$, even when
restricted to $\cipci$. This contrasts sharply with the globally hyperbolic
case,
where each timelike Killing field defines a unique ground state, and
different timelike Killing fields almost always give rise to the same
``frequency-splitting'' ground state~\cite{chmielowski}. 

\section{Conclusions}
\label{sec:conclusions}
The present work shows that a dynamical prescription with sensible
properties can be given in a large class of stationary spacetimes. In
the case of static spacetimes obeying \ref{cond:beta}, the present
prescription agrees with the previously studied second-order
prescription for a particular choice of extension, namely $A_F.$ The
basic prescription can also be modified to include all reasonable
dynamics (in the sense of \cite{nghsdII}) handled by the second-order
formalism. Nonetheless, the present work leaves many questions
unanswered.

First and foremost among these is whether the prescriptions given by
using the first-order formalism with different slicings in the same
spacetime give rise to the same dynamics.  While it is easy to
construct examples of spacetimes with inequivalent slicings satisfying
\ref{cond:beta}\footnote{For example, one may consider a Minkowski
``strip'' spacetime $(\mathbb{R}\times (0,1)^3, -dt^2 + \sum_{i=1}^3
dx_i^2),$ where the static slices are ``rotated into the time
direction.''}, it is not clear  whether the dynamics produced
by these slicings agree. It seems sensible that the preferred extension
$h^{SA}=T^{-1}$ would give rise to the same solutions in any slicing, but this
remains to be shown.  A related question is whether $h$ ever has
skew-adjoint extensions other than $T^{-1}$. The proof presented above
does not preclude this possibility, but no examples have thus far been
found, either. A different question is whether $h$ possesses a
skew-adjoint extension in an arbitrary stationary
spacetime. Assumptions (A) and (B) were used in crucial ways in the
proofs in Section \ref{sec:properties}. However, it is not known whether
Theorems 4.1 or 4.2 fail without these assumptions, or whether (A) and (B) are
only needed  because of the method of proof used. Along these same
lines, it would be interesting to compare the first and second-order
formalisms in static spacetimes when \ref{cond:beta} is dropped. If they could be shown to
agree, then the results of Theorem \ref{thm:main} would follow from the
proof in the second-order formalism, without imposing \ref{cond:beta}
(and possibly \ref{cond:mass}). On the other hand, when
\ref{cond:beta} fails, it is no longer true that $\X$-norm bounds some
multiple of $A^{\frac{1}{2}}$-norm. Hence $\X$ may properly contain
$q_F$, which considerably complicates the analysis of \mbox{Section
\ref{sec:static}.}

All of these question deal purely with the classical theory, but there
are questions regarding the quantum theory as well. The analysis of
Section \ref{sec:quantum} analyzed the linear theory only. In the past
few years, a general prescription for defining interacting quantum
field theory in globally hyperbolic, curved spacetimes has been given
\cite{Brunetti:1995rf,Brunetti:1999jn,Hollands:2001nf,Hollands:2001fb}.
This prescription relies on detailed properties of the so-called
``wavefront set'' of the Green's functions of the free theory. One
additional question to be answered is whether the solutions defined by the
prescription given in the present work allow the definition of Green's
functions satisfying these same conditions. This would allow an
interacting quantum theory to be defined in non-globally-hyperbolic
spacetimes in an analogous fashion.

\subsection*{Acknowledgments}
The author is greatly indebted to Robert Wald for his constant support
and advice. Thanks are also due to Stefan Hollands, Daniel
Hoyt, and Mike Seifert for many useful discussions. This work was
supported in part by a National Science Foundation Graduate Research
Fellowship and NSF Grant PHY00-90138 to the University of Chicago.

\end{document}